# INTERSTELLAR TITANIUM IN THE GALACTIC HALO

Keith Lipman[1] and Max Pettini[2,3]

[1] Institute of Astronomy, Madingley Road, Cambridge, CB3 0HA, UK
Electronic mail: kl@mail.ast.cam.ac.uk

[2] Royal Greenwich Observatory, Madingley Road, Cambridge, CB3 0EZ, UK
Electronic mail: pettini@mail.ast.cam.ac.uk

[3] Anglo-Australian Observatory, PO Box 296, Epping, NSW 2121, Australia

astro-ph/9410074  24 Oct 1994

## ABSTRACT

We present observations of Ti II $\lambda 3384$ absorption towards 15 distant stars in the Galactic halo and the Magellanic Clouds. These new data extend existing surveys of the distribution of $Ti^+$ to larger distances from the plane of the Galaxy than sampled previously, allowing the scale height of titanium to be determined for the first time. We find $h_{Ti^+} = 1.5 \pm 0.2$ kpc, a value which although greater than those of other tracers of neutral gas, is not as large as had been suspected. We interpret the extended distribution of $Ti^+$ as an indication that its severe depletion in interstellar clouds in the disk is reduced at the lower densities prevailing in the halo. The data are consistent with a simple power-law dependence of the Ti abundance on the ambient density, with exponent $k \simeq -1$. If the model is correct, it implies that refractory elements like Ti are fully returned to the gas phase at distances beyond $\sim 1$ kpc from the plane of the Galaxy.

## 1. INTRODUCTION

The determination of the extent, distribution, and kinematics of the interstellar medium away from the Galactic plane is important both for understanding the physical processes which drive the gas to large distances from the disk (e.g. Houck & Bregman 1990) and for interpreting the absorption produced by more distant galactic haloes in the spectra of background QSOs (e.g. Bowen, Blades, & Pettini 1995). This topic has been an active area of research over the last fifteen years, but the overall picture which has emerged is still incomplete. In particular, it is not yet clear how to bring together the information provided by different tracers of the ISM, which can exhibit significant differences in their distributions.

Highly ionized gas, discovered from C IV and Si IV absorption with *IUE* (Savage & de Boer 1979, Pettini & West 1982), has been shown by *HST* observations of extragalactic sources to have an exponential scale height $h \simeq 5$ kpc (Savage et al. 1993; see also the review by Sembach & Savage 1992). The patchiness of this gas, however, is such that the observations do not provide tight constraints on its large-scale distribution in the Galaxy; a spherical halo fits the data just as adequately as the more generally assumed exponential fall-off with distance from the plane, $| z |$.

Even the distribution of neutral hydrogen is open to different interpretations, despite the wealth of data collected with the *IUE* satellite and recently re-analyzed by Diplas & Savage (1994a). By considering the values of hydrogen column density, $N(H^0)$, for 374 sight-lines through the Galaxy, these authors deduced a scale height $h_{H^0} \simeq 0.2$ kpc, although the observations are equally well represented by the sum of *two* exponential distributions, with $h_1 = 0.12$ kpc and $h_2 = 1.2$ kpc respectively. That significant amounts of $H^0$ do exist at $| z |$ distances greater than 1 kpc is demonstrated by the comparison between $N(H^0)$ measures from Lyman $\alpha$ absorption towards halo stars and from 21 cm emission, the latter sampling the full extent of the gas in a given direction, including material beyond the stars (Lockman, Hobbs, & Shull 1986; Danly et al. 1992).

In disk clouds, refractory elements are mostly removed from the gas phase and incorporated into dust grains. Edgar & Savage (1989) showed that the scale heights of species such as $Ti^+$, $Ca^+$ and $Fe^+$ are significantly larger than that of $H^0$, in broad agreement with earlier findings that the depletions of these elements exhibit a dependence on $< n(H^0) >$, the average gas volume density along a line of sight (e.g. Savage & Bohlin 1979; Phillips, Pettini, & Gondhalekar 1984). Presumably, as $n(H^0)$ decreases with distance from the plane of the Galaxy, the reduced depletions of Ti, Ca and Fe compensate for the overall decrease in gas density, leading to shallower profiles in $| z |$ for these species than for neutral hydrogen. Spitzer (1985) and Jenkins, Savage, & Spitzer (1986) proposed that the correlation of depletion with mean line-of-sight density can be understood in terms of a simple model of the ISM consisting of diffuse cold clouds and of a warm intercloud gas. In this picture the refractory elements are more severely depleted in the clouds than in the intercloud medium; the observed correlations follow from the fact that $< n(H^0) >$ is a rough measure of the relative proportions of the two phases along a given line of sight. The two components may well have different scale heights, with the lower density intercloud gas conceivably extending further into the halo than the cold clouds. Species such as $Ca^+$ and $Ti^+$ are then useful tracers of a diffuse interstellar medium which may be found out to large distances from the disk.

$Ti^+$ is especially favoured in this context. Its ionization potential (13.58 eV) is so close to that of $H^0$ that concerns about unobserved ion stages are minimized and the Ti abundance in the neutral gas can be deduced directly. Furthermore, the strongest transition of Ti II, at $\lambda 3383.768$ with an $f$-value of 0.3401 (Morton 1991), is normally unsaturated, allowing a reliable measurement of the column density from the line equivalent width. These properties contrast with those of other interstellar species, such as $Ca^+$ and $Na^0$, accessible to ground-based telescopes. On the other hand the line occurs in the near-ultraviolet, where atmospheric extinction is high and in general the sensitivity of optical instrumentation is low. Therefore, observations of Ti II absorption in the Galactic halo have until recently been limited to the brightest stars and relatively little data are available for distances beyond 1 kpc.

In particular, there are hardly any published measurements of Ti II in the spectra of extragalactic sources, allowing the *total* column density perpendicular to the Galactic plane, $N_0(Ti^+)$, to be deduced. Consequently, previous attempts to determine the scale height of $Ti^+$ have been inconclusive. Edgar & Savage (1989) reported a lower limit $h_{Ti^+} \geq 2$ kpc, while $h_{Ca^+} = 1$ kpc. By adding new observations of 25 sight-lines to earlier compilations in a comprehensive study of Ti II and Ca II in the halo, Albert et al. (1993) confirmed the scale height of $Ca^+$, but were still unable to see a turn-over in the vertical distribution of $Ti^+$.

In this paper we remedy this situation by reporting observations of Ti II absorption in the spectra of 15 stars located beyond 1 kpc from the plane of the Galaxy and including four stars in the Magellanic Clouds. These data extend previous studies of the distribution of $Ti^+$ in the halo to a largely unexplored region and, as we discuss below, show that the scale height of $Ti^+$ is in fact $h_{Ti^+} \simeq 1.5$ kpc, greater than that of $Ca^+$, but not by a large factor.

## 2. OBSERVATIONS AND DATA REDUCTION

Table 1 lists relevant properties of the 15 stars observed in this study, including their spectral types and distances from the Sun. Nominally, the distance estimates are accurate to about $\pm 25\%$, but the question of whether these stars have been correctly classified is still open. Although the presence of young B stars several kpc from the plane is surprising, the work of Keenan and collaborators in particular has shown that many of these stars have temperatures, gravities and *chemical abundances* indistinguishable from their disk counterparts and are therefore probably *bona fide* Population I stars (Tobin 1991 and Keenan 1992 give recent discussions of this problem). In the present work we have adopted, whenever possible, distances deduced by fine analyses of high dispersion stellar spectra, such as those published by the Belfast group (see Little et al. 1994 for a description of the method). For the stars in Magellanic Clouds we have adopted distances of 50 kpc to the LMC and 57 kpc to the SMC from the review by Feast (1991).

The data reported here were secured over a period of 5 years, from 1989 to 1994. All but one of the stars were observed with the University College London echelle spectrograph (UCLES) at the coudé focus of the 3.9 m Anglo-Australian Telescope at Siding Spring Observatory, Australia. The observations were mostly carried out in periods of poor observing conditions, as back-up to a more demanding programme of faint-object spectroscopy which has been described elsewhere (Pettini et al. 1990). The spectrum of one star, HDE 233622, was obtained with the Utrecht echelle spectrograph (UES) at the Nasmyth focus of the 4.2 m William Herschel Telescope on La Palma, Canary Islands; this instrument is very similar to UCLES. With both UCLES and UES we used the 79 grooves $mm^{-1}$ echelle gratings, which result in an inter-order separation in excess of 20 arcsec at the wavelength of interest here, thereby providing adequate sky signal for accurate background subtraction. All the spectra—except those of HD 5980 and HD 269333— were recorded with the IPCS, a photon counting detector which, because of its low noise, high quantum efficiency in the ultraviolet and small pixels, is well suited to the present programme. HD 5980 and HD 269333 were observed in 1994 with a $1024 \times 1024$ Tektronix CCD. The slit width was set to 0.8 arcsec for the IPCS observations and increased to 1 arcsec when using the CCD which has larger pixels.

The two-dimensional echelle spectra were reduced with standard procedures using the STARLINK software packages FIGARO, DIPSO and APIG; a detailed description of the data reduction steps is given by Pettini et al. (1990). In general, several exposures were obtained for each star in Table 1. Individual spectra, wavelength calibrated and sky-subtracted, were rebinned to a common Local Standard of Rest (LSR) wavelength scale before being added together. The profiles of emission lines from the Th-Ar hollow cathode lamp used for wavelength comparison show the resolution of the final spectra to be FWHM $= 0.060 \pm 0.003$ Å, sampled with $\sim 2.4$ IPCS pixels; at the wavelength of Ti II $\lambda 3383.768$, the corresponding velocity resolution is $5.3 \pm 0.3$ km $s^{-1}$. The resolution of the two CCD spectra is 25% lower (FWHM $= 6.6$ km $s^{-1}$).

TABLE 1

**The Stars Observed**

| Star | $l$ | $b$ | R.A. (1950) | Dec. (1950) | Spectral Type | V | B−V | E(B−V) | d (kpc) | z (kpc) | $v \sin i$ (km s$^{-1}$) | Refs. | Date Observed |
|------|-----|-----|-------------|-------------|---------------|---|-----|--------|---------|---------|------------|-------|---------------|
| HD 5980 | 302.1 | −44.9 | 00 57 22.9 | −72 26 36 | WP + OB + Neb: | 11.81 | −0.22 | 0.09 | 57[a] | −40[a] | ... | 1,2,3 | 21 Aug 1994 |
| HD 121968 | 334.0 | 55.8 | 13 56 15.8 | −02 40 19.9 | B1 V | 10.31 | −0.19 | 0.07 | 5.2 | 4.3 | >150 | 4 | 12 & 13 Aug 1991 |
| HD 125924 | 338.2 | 48.3 | 14 20 03.77 | −08 01 15.9 | B2 IV | 9.68 | −0.19 | 0.05 | 3.2 | 2.4 | 105 | 4 | 11 Aug 1991 |
| HD 158243 | 337.6 | −10.6 | 17 27 03.97 | −53 26 28.1 | B1 Iab | 8.15 | 0.00 | 0.19 | 6.5 | −1.2 | ... | 5 | 14 Aug 1991 |
| HD 163522 | 349.6 | −9.1 | 17 55 00.23 | −42 28 55.9 | B1 Ia | 8.44 | 0.00 | 0.19 | 9.3 | −1.5 | ∼ 50 | 5,6 | 3 Jun 1990 |
| HD 179407 | 24.2 | −10.4 | 19 10 04.94 | −12 40 04.9 | B0.5 Ib | 9.30 | 0.06 | 0.28 | 5.4 | −1.0 | 156 | 7 | 10 & 13 Aug 1991 |
| HD 195455 | 20.3 | −32.1 | 20 29 17.22 | −24 24 15.7 | B0.5 III | 9.20 | −0.18 | 0.08 | 3.1 | −1.6 | 200 | 7 | 14 Aug 1991 |
| HD 206144 | 34.8 | −45.1 | 21 37 47.31 | −17 49 39.2 | B1 III | 9.41 | −0.15 | 0.11 | 5.0 | −3.6 | 64 | 7 | 10 Aug 1991 |
| HD 209684 | 43.5 | −49.1 | 22 02 49.36 | −14 00 49.8 | B3 V | 9.86 | −0.20 | 0.00 | 1.7 | −1.3 | ... | 4 | 26 Aug 1990 & 14 Aug 1991 |
| HD 220787 | 67.8 | −64.4 | 23 24 09.74 | −11 18 27.2 | B3 III | 8.30 | −0.17 | 0.03 | 1.4 | −1.3 | 27 | 7 | 21 Jul 1989 |
| HDE 233622 | 168.2 | 44.2 | 09 18 08.26 | 50 18 43.9 | B1.5 | 9.97 | −0.23 | 0.02 | 3.5 | 2.4 | 240 | 4 | 5 Dec 1992 |
| HD 269006 | 282.8 | −34.2 | 05 02 44.2 | −71 24 15 | Peculiar | 9.83 | ... | ... | 50[b] | −26[b] | ... | 1,3 | 14 Aug 1991 |
| HD 269333 | 279.9 | −33.4 | 05 18 39.2 | −69 14 45 | W + B1:I | 11.21 | −0.09 | 0.11 | 50[b] | −26[b] | ... | 1,8,3 | 21 Aug 1994 |
| JL 212 | 303.6 | −61.0 | 00 46 48.35 | −56 22 09.4 | B2 | 10.32 | −0.20 | 0.04 | 3.6 | −3.2 | 219 | 7 | 26 Aug 1990 |
| R 136 | 279.4 | −31.7 | 05 39 04.4 | −69 07 35 | OB(n)+ WN5−A(B) | 10.77 | 0.14 | 0.34 | 50[b] | −26[b] | ... | 9,3 | 24 Feb 1990 |

[a] In the Small Magellanic Cloud

[b] In the Large Magellanic Cloud

**TABLE 2**
**Equivalent Widths and Column Densities**

| Star | $b$ | $z$ (kpc) | S/N | $W_\lambda(3384)$ (mÅ) | $N(\mathrm{Ti}^+)$ $(10^{11}\mathrm{cm}^{-2})$ | $N(\mathrm{H}^0)$ [1] $(10^{20}\mathrm{cm}^{-2})$ |
|------|-----|-----------|-----|------------------------|-------------------------------------------------|----------------------------------------------------|
| HD 5980 | $-44.9$ | $-40$ | 12 | $41 \pm 11$ | $13.0 \pm 3.5$ | 1.9 |
| HD 121968 | 55.8 | 4.3 | 34 | $38 \pm 2$ | $12.0 \pm 0.8$ | $4.0 \pm 0.8$ |
| HD 125924 | 48.3 | 2.4 | 28 | $46 \pm 5$ | $14.3 \pm 1.6$ | $4.6 \pm 0.9$ |
| HD 158243 | $-10.6$ | $-1.2$ | 30 | $91 \pm 4$ | $30.7 \pm 1.4$ | $12.9 \pm 3.0$ |
| HD 163522 | $-9.1$ | $-1.5$ | 57 | $178 \pm 3$ | $60.4 \pm 2.7$ | $12.3 \pm 2.3$ |
| HD 179407 | $-10.4$ | $-1.0$ | 35 | $83 \pm 4$ | $27.2 \pm 1.3$ | $12.9 \pm 3.2$ |
| HD 195455 | $-32.1$ | $-1.6$ | 35 | $49 \pm 3$ | $15.3 \pm 0.9$ | $4.7 \pm 1.1$ |
| HD 206144 | $-45.1$ | $-3.6$ | 31 | $51 \pm 4$ | $16.6 \pm 1.3$ | $3.2 \pm 1.0$ |
| HD 209684 | $-49.1$ | $-1.3$ | 33 | $51 \pm 4$ | $16.2 \pm 1.3$ | $\ldots$ |
| HD 220787 | $-64.4$ | $-1.3$ | 40 | $26 \pm 3$ | $8.0 \pm 0.9$ | $\ldots$ |
| HDE 233622 | 44.2 | 2.4 | 55 | $27 \pm 3$ | $8.1 \pm 0.9$ | $\ldots$ |
| HD 269006 | $-34.2$ | $-26$ | 12 | $55 \pm 12$ | $17.3 \pm 3.8$ | $\ldots$ |
| HD 269333 | $-33.4$ | $-26$ | 20 | $28 \pm 4$ | $9.3 \pm 1.4$ | 4.3 |
| JL 212 | $-61.0$ | $-3.2$ | 22 | $42 \pm 5$ | $13.0 \pm 1.5$ | $\ldots$ |
| R 136 | $-31.7$ | $-26$ | 26 | $62 \pm 7$ | $21.5 \pm 2.4$ | 3.2 |

[1] H I data from Lyman $\alpha$ measurements by Diplas & Savage (1993a) except for HD 5980, HD 269333 and R 136 which are from the 21cm observations by McGee, Newton & Morton 1983.

Portions of the echelle spectra encompassing the interstellar Ti II line are reproduced on an expanded vertical scale in Figure 1; the spectra have been normalized to the local stellar continuum. As can be seen from the Figure, although interstellar Ti II absorption is detected in every case, the signal-to-noise ratio varies significantly from star to star. Values of S/N, deduced from the rms deviation of the data points from the fitted stellar continuum, are listed in column 4 of Table 2. Typically, S/N $\approx 30$; the corresponding $3\sigma$ detection limit for the equivalent width of an unresolved absorption line is $W_\lambda(3\sigma) \simeq 6$ mÅ. The measured equivalent widths of the Ti II lines are listed in column 5 of Table 2, together with $1\sigma$ errors which reflect only the counting statistics (and do not include uncertainties in the continuum placement). From Figure 1 it can be seen that in the case of R 136, and possibly HD 269006, we detect Ti II absorption at LMC velocities of between $v_{LSR} \simeq 200$ and $300$ km s$^{-1}$; in HD 5980 there is a $3.5\sigma$ absorption feature at SMC velocities ($\approx 140$ km s$^{-1}$).

## HD 5980

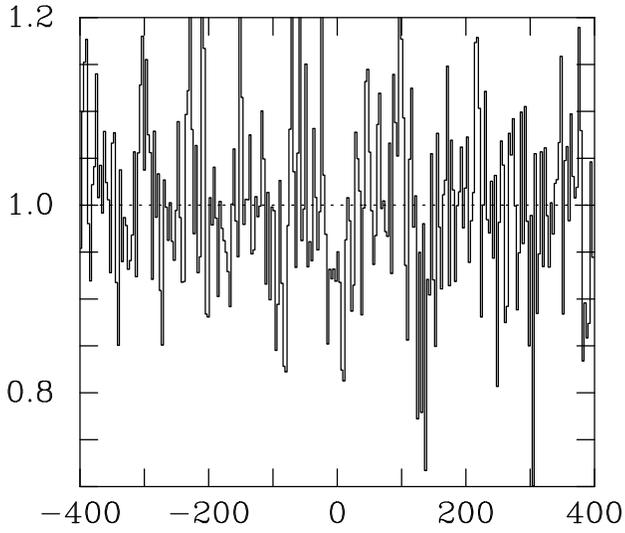

## HD 121968

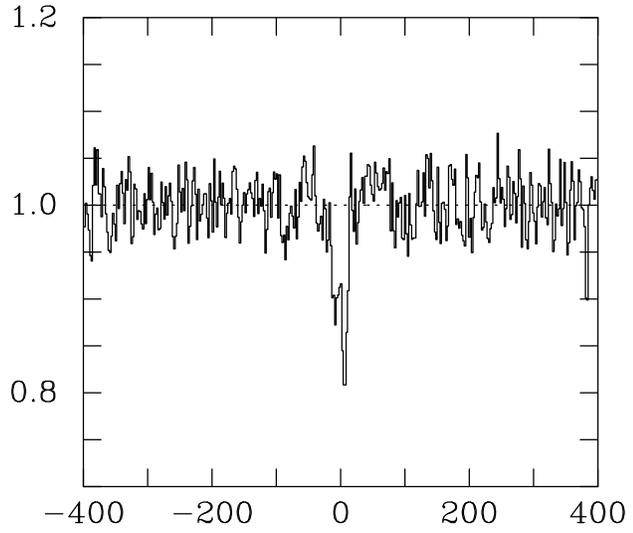

## HD 125924

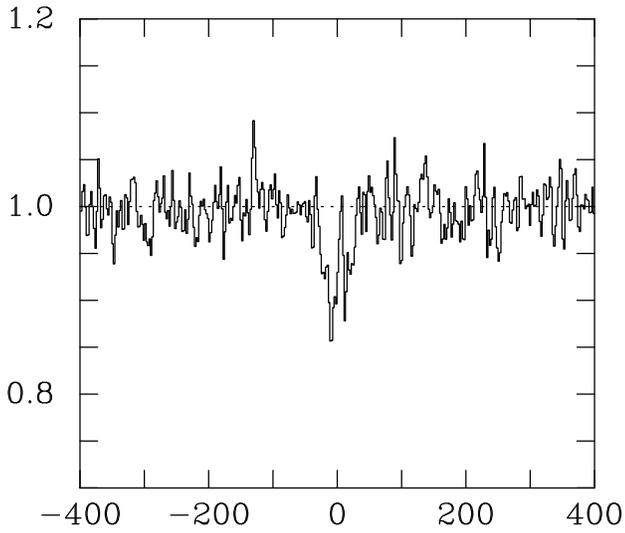

## HD 158243

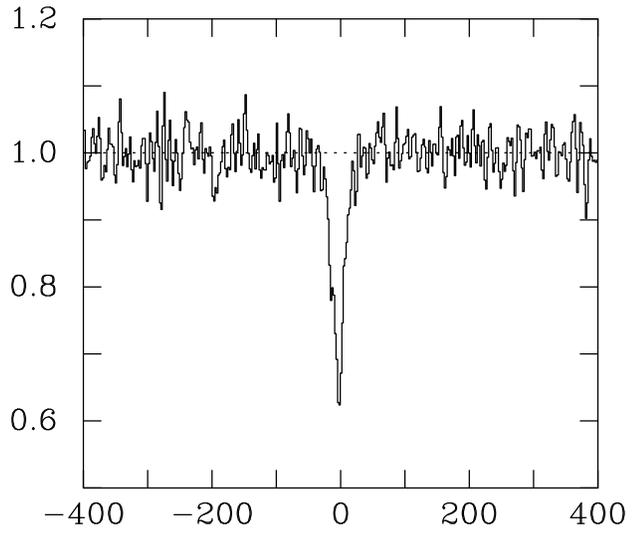

## HD 163522

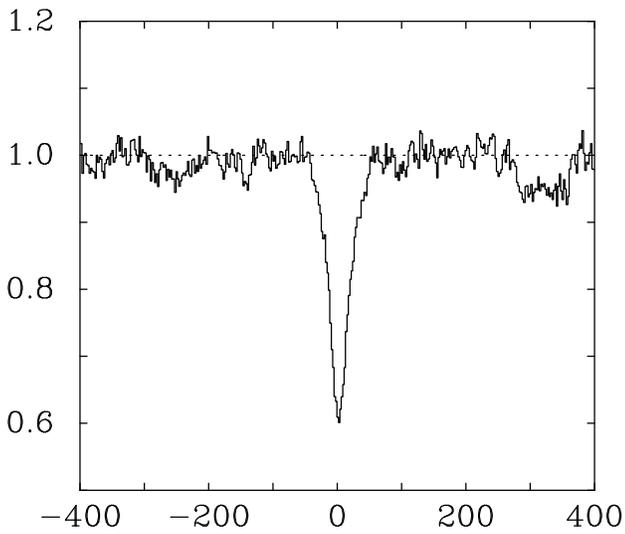

## HD 179407

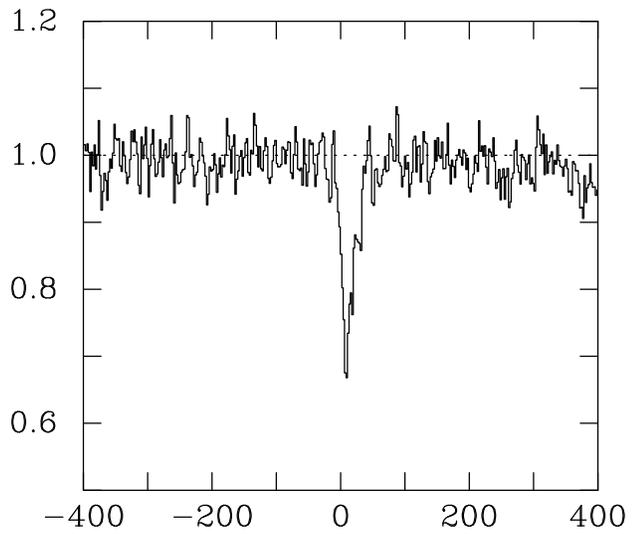

## HD 195455

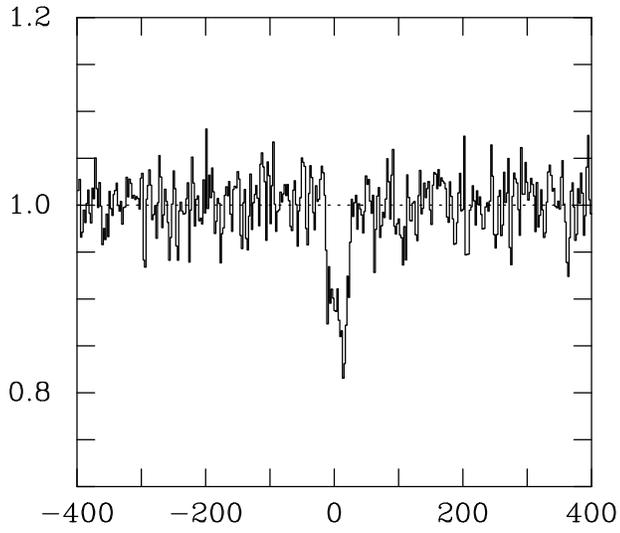

## HD 206144

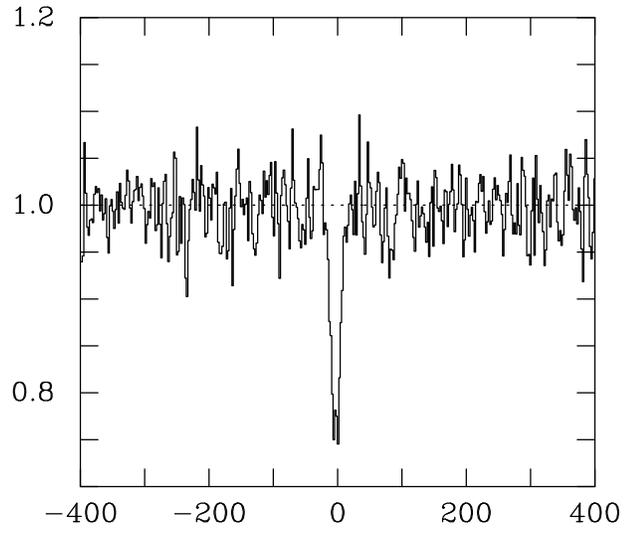

## HD 209684

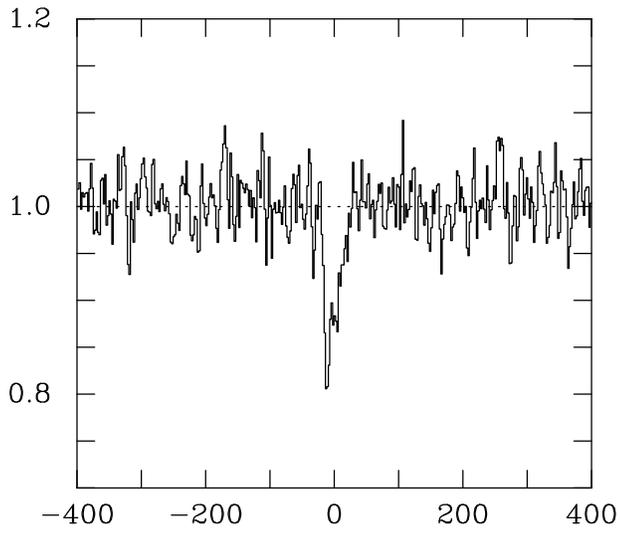

## HD 220787

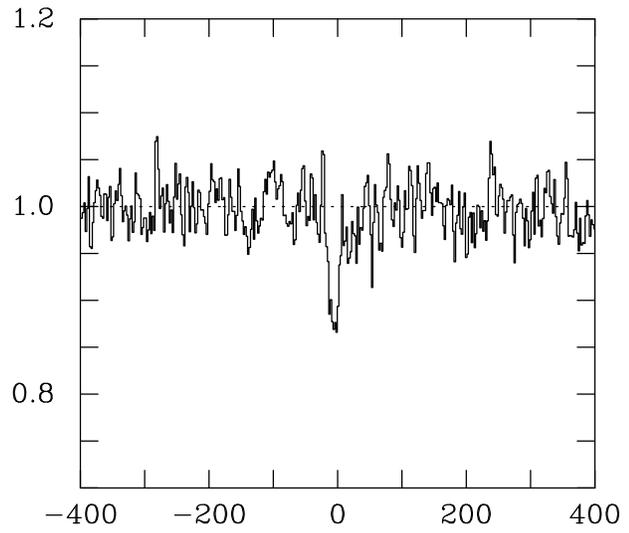

## HDE 233622

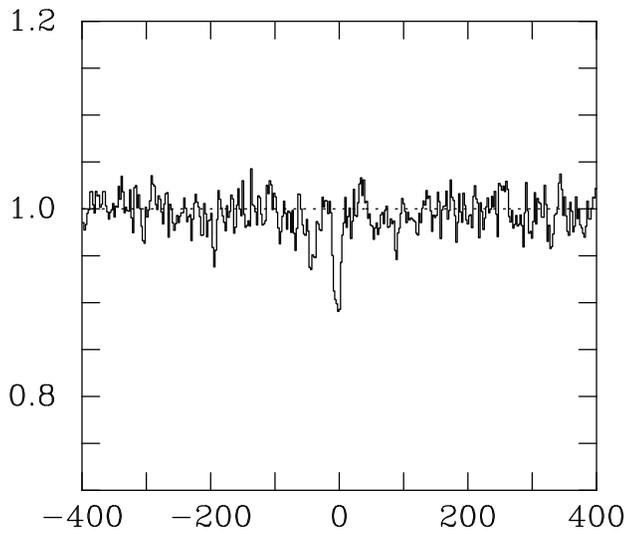

## HD 269006

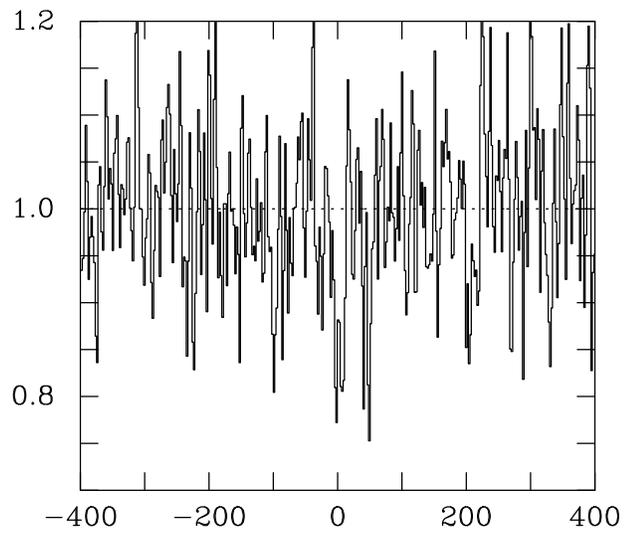

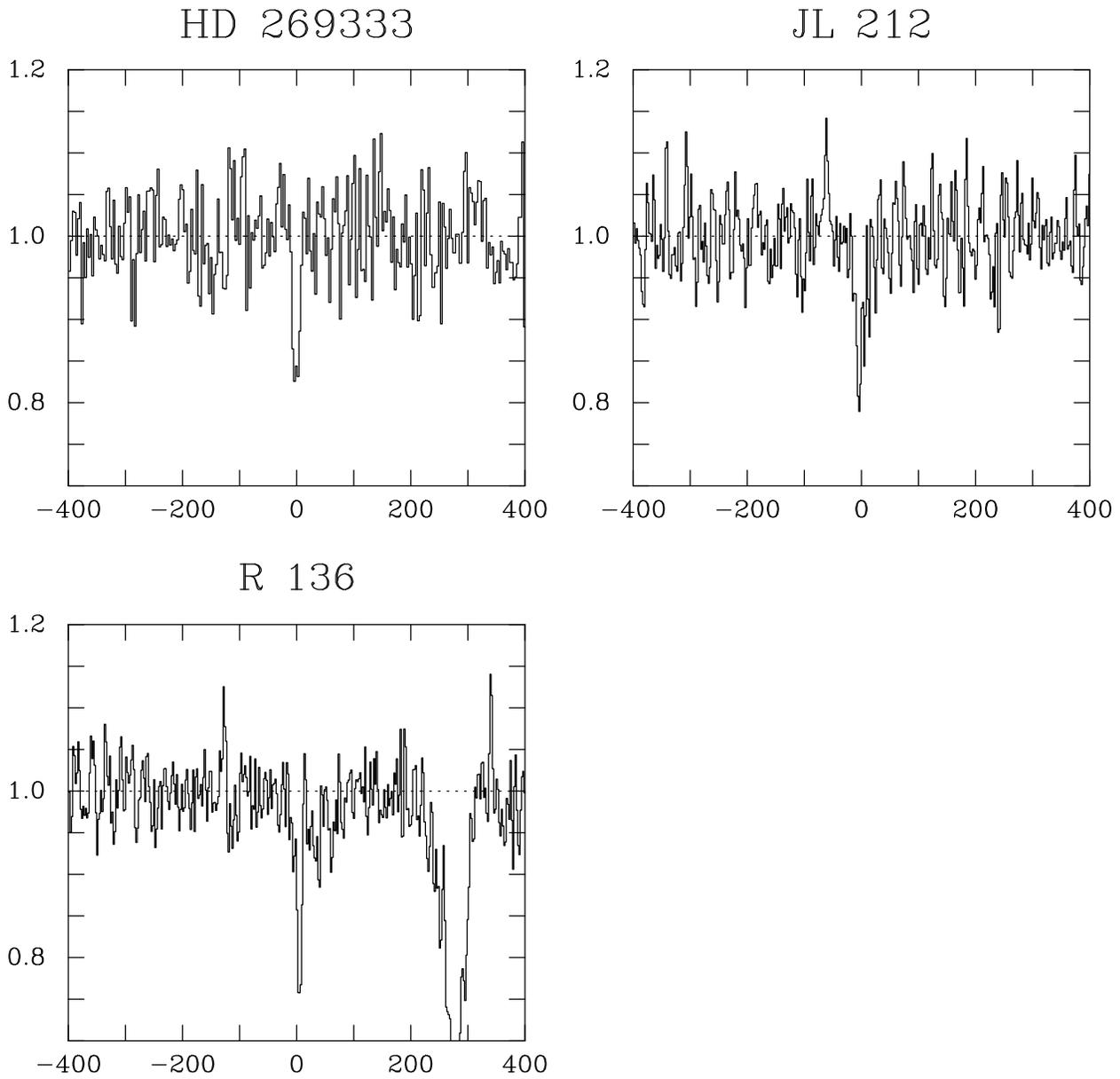

Fig. 1.— Normalized profiles of Ti II λ3383.768 absorption lines towards 15 distant stars, including one in the Small Magellanic Cloud (HD 5980) and three in the Large Magellanic Cloud (HD 269006, HD 269333, and R 136). The x-axis gives the velocity in km s$^{-1}$ relative to the Local Standard of Rest; the y-axis shows the residual intensity on an expanded scale.

Finally, in column 6 of Table 2 we give values of column density $N(\mathrm{Ti}^+)$, deduced by fitting theoretical profiles to the observed absorption lines (Pettini et al. 1990). Details of the profile fits ($v_{\mathrm{LSR}}$, velocity dispersion parameter $b$, and $N(\mathrm{Ti}^+)$) are collected in Table 3. Since the Ti II lines are fully resolved and generally weak, we do not expect significant problems from unrecognized, saturated components which, if present, could lead to an underestimate of the total column density (Nachman & Hobbs 1973). The errors listed in column 6 of Table 2 reflect only the quoted uncertainties in equivalent width.

Some of our echelle spectra also encompass a resonance line of neutral titanium, Ti I $\lambda 3341.873$, with $f = 0.1517$ (Morton 1991). As expected, this line is not detected since $\mathrm{Ti}^0$ is a minor ionization stage in H I regions; we can place a typical $3\sigma$ upper limit of $\sim 25\%$ to the contribution by $\mathrm{Ti}^0$ to the total column density of titanium in the H I gas sampled by our observations.

## 3. THE SCALE HEIGHT OF INTERSTELLAR TITANIUM

In order to study the distribution of titanium in the Galactic halo we have combined our measurements with those published in the extensive compilations by Edgar & Savage (1989) and by Albert et al. (1993). The values of $N(\mathrm{Ti}^+)$ in the compilation by Edgar & Savage have been multiplied by a factor of 0.957 to take into account the different $f$-value of Ti II $\lambda 3384$ adopted by these authors. In subsequent figures we have used different symbols to indicate the three sources of data, as follows: Edgar & Savage — circles; Albert et al. — squares; this paper — triangles. The combined sample consists of a total of 116 sight-lines; the new observations presented here increase the number of objects further than 2 kpc from the Galactic plane from 3 to 12.

Only one of our 15 stars, HD 220787, is in common with earlier work. Albert et al. (1993) report $W_\lambda(\lambda 3384) = 15 \pm 5$ mÅ, compared with $26 \pm 3$ mÅ measured from our spectrum. Accordingly, the value of $N(\mathrm{Ti}^+)$ we deduce is $\sim 70\%$ higher (Table 2). As this work was being completed we were kindly sent a preprint of the recent paper by Albert, Welsh, & Danly (1994) reporting Ti II observations towards six halo stars two of which, HD 195455 and HD 206144, are in common with our sample. Again the agreement is not good, with our values of $W_\lambda(\lambda 3384)$ being $\sim 80\%$ higher. These discrepancies are a concern, but we suspect that they may be due to the lower signal-to-noise ratio of the spectra by Albert et al. (1993 and 1994). In the present analysis we have adopted our own measures for the three stars in question.

Figure 2 shows the increase in the column density of $\mathrm{Ti}^+$, projected along a direction perpendicular to the Galactic plane, with distance from the plane $|\,z\,|$ for the full sample. It is worth noting, in passing, that by considering all the data together, we make the tacit assumptions that there are no significant differences in the distribution of titanium above and below the plane, nor any trends with Galactic longitude. With a moderately large sample, such as the one under consideration here, it is possible to check the validity of these assumptions. Indeed, we find that the two sub-samples, at positive and negative Galactic latitudes respectively, are statistically indistinguishable, nor do we see any trends in the abundance of titanium with distance from the centre of the Galaxy. Presumably the range of distances probed is too small to show the effects of radial abundance gradients (see also Savage & Massa 1985).

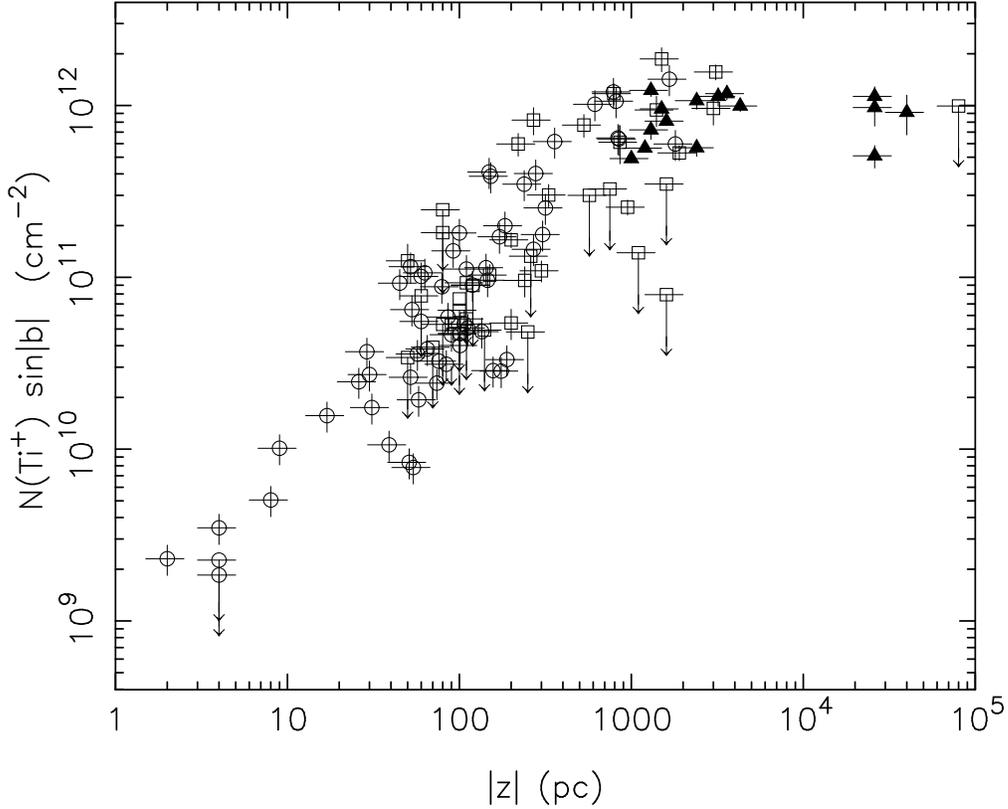

Fig. 2. — Column density of $Ti^+$, projected along the $z$–direction, plotted as a function of distance from the plane of the Galaxy. The total sample consists of 116 measurements (or upper limits); different symbols denote different compilations of data, as follows: Edgar & Savage (1989) — circles; Albert et al. (1993) — squares; this paper — filled triangles. The errors in $N(Ti^+)$ are those reported by the original authors; the uncertainties in the stellar distances are a nominal $\pm 25\%$. The upper limit at the largest distance refers to the sight-line to the Seyfert 2 galaxy NGC 1068; for convenience this point is plotted at $|z| = 80$ kpc. The increase in $N(Ti^+)$ with $|z|$ appears to level off between $|z| \simeq 1$ and 3 kpc.

Returning to Figure 2, we see that the increase in $N(\mathrm{Ti^+})$ with $|z|$ known from earlier studies appears to level off between 1 and 3 kpc from the plane, and that the values of $N(\mathrm{Ti^+})$ measured towards the Magellanic Clouds—as well as the upper limit reported by Blades & Morton (1983) towards NGC 1068—are no larger than those for distant stars in the halo. If we assume an exponential distribution of the form:

$$n_z(\mathrm{Ti^+}) = n_0(\mathrm{Ti^+}) \exp\left[-|z|/h_{Ti^+}\right] \qquad (1)$$

the column density of $\mathrm{Ti^+}$ in front of star at height $|z|$ is given by:

$$N \sin|b| = N_0(1 - \exp\left[-|z|/h_{Ti^+}\right]) \qquad (2)$$

where

$$N_0 = n_0(\mathrm{Ti^+}) \times h_{Ti^+} \qquad (3)$$

and $b$ is the Galactic latitude.

Fitting eq. 2 to the data in Figure 2 and minimizing $\chi^2$ in log-log space yielded

$$N_0(\mathrm{Ti^+}) = 1.3 \times 10^{12} \ \mathrm{cm^{-2}} \ \text{and} \ h_{Ti^+} = 1.5 \ \mathrm{kpc} \qquad (4)$$

For some of the 116 sight-lines considered, only upper limits to $N(\mathrm{Ti^+})$ are available. We treated such upper limits as detections, at a value of $N(\mathrm{Ti^+})$ a factor of 2 lower than the upper limit, and assumed an uncertainty of a factor of 2 (that is, significantly larger than the typical errors in $N(\mathrm{Ti^+})$ in cases where the line is detected—see Figure 2). While this procedure is not mathematically rigorous, we feel that it is a better approximation than the common approach of simply ignoring the non-detections.

The above fitting procedure also showed that eq. 1 is not an adequate representation of the distribution of titanium in the $z-$direction, since $\chi^2/\nu = 14.9$, where $\nu$ is the number of degrees of freedom. This was not unexpected, since earlier analyses of this kind have reached the same conclusion. It is normally assumed that the large scatter in plots such as Figure 2 is due to the intrinsic 'patchiness' of the ISM on the scales sampled, although errors in the stellar distances presumably also contribute. We followed Edgar & Savage (1989) in introducing a random scattering factor, $\sigma_{\mathrm{rdm}}$, which was added in quadrature to the individual errors in $N(\mathrm{Ti^+})$. We then solved for $N_0(\mathrm{Ti^+})$ and $h_{Ti^+}$, forcing $\chi^2/\nu$ to 1. Table 4 shows the effect on $N_0(\mathrm{Ti^+})$ and $h_{Ti^+}$ when $\sigma_{\mathrm{rdm}}$ is increased from 0 to 0.30 dex; $\chi^2/\nu$ reaches unity when $\sigma_{\mathrm{rdm}} = 0.29$ dex, or about a factor of 2. The corresponding fit

$$N_0(\mathrm{Ti^+}) = 1.1 \times 10^{12} \ \mathrm{cm^{-2}} \ \text{and} \ h_{Ti^+} = 1.35 \ \mathrm{kpc} \qquad (5)$$

compared to the values for $\sigma_{\mathrm{rdm}} = 0.00$ dex (eq. 4) gives an indication of the uncertainties in these quantities. We conclude that, while $h_{Ti^+}$ is apparently somewhat greater than the scale height of $\mathrm{Ca^+}$ ($h_{Ca^+} = 1$ kpc; Edgar & Savage 1989), the present data do not support earlier

**TABLE 4**

**Exponential Distribution Models**

| $\sigma_{rdm}$ | All Data (116 sightlines) | | | Galactic Data (111 sightlines) | | |
|---|---|---|---|---|---|---|
| (dex) | $\log[N_0(\mathrm{Ti}^+)]$ (cm$^{-2}$) | $h_{Ti^+}$ (pc) | $\chi^2/\nu$ | $\log[N_0(\mathrm{Ti}^+)]$ (cm$^{-2}$) | $h_{Ti^+}$ (pc) | $\chi^2/\nu$ |
| 0.00 | 12.12 | 1533 | 14.94 | 12.16 | 1770 | 15.09 |
| 0.10 | 12.06 | 1404 | 5.12 | 12.15 | 1753 | 5.23 |
| 0.20 | 12.04 | 1378 | 1.86 | 12.13 | 1702 | 1.91 |
| 0.30 | 12.03 | 1342 | 0.94 | 12.10 | 1604 | 0.97 |

suggestions that the layer of Ti$^+$ in the Galaxy may be several kpc thick. In particular, we note that Albert et al. (1994) proposed such an extended distribution of Ti$^+$ largely on the basis of their observations of *one* star, HD 123884, of spectral type B8/9Ia/abp apparently located 8.7 kpc above the Galactic plane. Albert et al. found Ti II absorption to be exceedingly strong in this direction; $W_\lambda(\lambda 3384) = 172$ mÅ corresponding to $N(\mathrm{Ti}^+)\sin|b| = 3.9 \times 10^{12}$ cm$^{-2}$, larger than any of the values in our Figure 2. Setting aside the uncertainty in the spectral classification—and therefore distance—of the star, the fact that we measure significantly lower values of $N(\mathrm{Ti}^+)$ along sightlines to extragalactic sources suggests that either the column density of titanium is unusually high towards HD 123884 or the absorption feature identified by Albert et al. is partly of stellar, rather than interstellar, origin.

Returning to our Figure 2, the intrinsic scatter in the distribution of Ti$^+$, $\sigma_{rdm} = 0.3$ dex, is lower than that of the $N(\mathrm{H}^0)$ distribution in the same stars (see §4.1 below), $\sigma_{rdm}(\mathrm{H}^0) = 0.45$ dex, consistent with the claim by Edgar & Savage that Ti$^+$ is distributed more smoothly than the general neutral ISM.

Finally, we point out that the turn-over in the distribution of Ti$^+$ seen in Figure 2 does *not* hinge on the (few) observations of extragalactic sources available at present. If we repeat the analysis excluding the four stars in the Magellanic Clouds and the upper limit for NGC 1068, we deduce

$$N_0(\mathrm{Ti}^+) = 1.4 \times 10^{12} \text{ cm}^{-2} \text{ and } h_{Ti^+} = 1.8 \text{ kpc} \qquad (6)$$

which are only slightly larger than the values appropriate to the full sample (see Table 4). While additional observations along sight-lines out of the Galaxy would obviously be very valuable, the levelling off of $N(\mathrm{Ti}^+)$ between $|z| = 1 - 3$ kpc is already apparent in the spectra of distant halo stars.

## 4. THE DEPLETION OF TITANIUM IN THE GALACTIC HALO

The scale height of Ti$^+$ is the largest measured so far for a dominant ion stage in H I gas. As discussed in the Introduction, this result is generally interpreted as being due to the reduced depletion of refractory elements at the low densities which are presumably found away from the disk of the Galaxy. In order to investigate this effect quantitatively, it is necessary to know the column density of neutral gas, $N(\mathrm{H}^0)$, along the lines of sight studied.

### 4.1 *Neutral Hydrogen Column Densities*

Values of $N(\mathrm{H}^0)$ are available for 107 out of the 116 directions considered. For nearly all of the stars in the compilation by Edgar & Savage (1989) $N(\mathrm{H}^0)$ has been deduced from observations of the Lyman $\alpha$ absorption line. On the other hand, the values of $N(\mathrm{H}^0)$ reported by Albert et al. (1993) are from 21 cm emission spectra with an angular resolution of 21 arcmin. These measurements may include material *beyond* the stars, and therefore not associated with the observed Ti II, and may also be affected by clumpiness of the ISM within the radio beam. However, the corrections required do not appear to be very large, at least for the six stars in the Albert et al. survey which have subsequently been measured at Lyman $\alpha$ by Diplas & Savage 1994b (see also Figure 5 of Danly et al. 1992). Specifically, for five of these stars (HD 60848, HD 119608, HD 120086, HD 149881 and HD 203664), $N(\mathrm{H}^0)_{Ly\alpha}$ is smaller than $N(\mathrm{H}^0)_{21\ cm}$ by between $\approx 5$ and $\approx 30\%$, while for the sixth (HD 214080) $N(\mathrm{H}^0)_{Ly\alpha}$ is $\approx 25\%$ *greater* than $N(\mathrm{H}^0)_{21\ cm}$ (this is because Albert et al. attempted to correct for H I beyond the stars by considering only 21 cm emission over the velocity range encompassed by the optical absorption lines). In all six cases we have adopted $N(\mathrm{H}^0)_{Ly\alpha}$. Lyman $\alpha$ absorption measurements are available for 7 of the 15 stars we have observed (see Table 2); for the Galactic sight-lines to 3 of the 4 stars in the Magellanic Clouds values of $N(\mathrm{H}^0)_{21\ cm}$ have been reported by McGee, Newton, & Morton (1983). Finally on this point, molecular hydrogen has been observed in only a subset of the 107 stars; however, the molecular fraction of gas in the halo is likely to be small (Savage et al. 1977).

One way to assess the global properties of our sample of 107 $N(\mathrm{H}^0)$ measurements is to determine the the vertical distribution of the neutral gas (in a plot of $N(\mathrm{H}^0){\times}\sin|b|$ vs. $|z|$ — not reproduced here) and compare it with that of the larger data set of 374 sight-lines considered by Diplas & Savage (1994a). The best fit to our hydrogen data assuming an exponential distribution is obtained for

$$N_0(\mathrm{H}^0) = 2.3 \times 10^{20}\ \mathrm{cm}^{-2} \ \mathrm{and}\ h_{H^0} = 95\ \mathrm{pc}\,. \tag{7}$$

This value of $N_0(\mathrm{H}^0)$ corresponds to a mean volume density in the disk $n_0(\mathrm{H}^0) = 0.78\ \mathrm{cm}^{-3}$; for comparison Diplas & Savage found $n_0(\mathrm{H}^0) = 0.366\ \mathrm{cm}^{-3}$ and $h_{H^0} = 195\ \mathrm{pc}\,$. Evidently the sample of stars for which Ti II measurements are available is less biased against large, dense interstellar clouds than the *IUE* H I survey by Diplas & Savage.

## 4.2 *Density-dependent Depletion of Titanium*

Figure 3 shows how the observed ratio of column densities of Ti$^+$ and H$^0$ varies with distance from the plane of the Galaxy. For comparison, the solar abundance is $A(\mathrm{Ti})_\odot = 8.5 \times 10^{-8}$ (Anders & Grevesse 1989), beyond the limit of the plot. The lines drawn are the predictions of a simple model of the depletion of Ti discussed below. From Figure 3 it appears that the large and highly variable depletion of Ti, well known from earlier surveys, is confined to the disk of the Galaxy. The median $A(\mathrm{Ti})_{ISM} = 4.2 \times 10^{-10}$ for stars with $|z| \leq 250$ pc corresponds to a logarithmic depletion $D(\mathrm{Ti}) = -2.3$, where

$$D(\mathrm{Ti}) = \log\left[A(\mathrm{Ti})_{ISM}/A(\mathrm{Ti})_\odot\right] \tag{8}$$

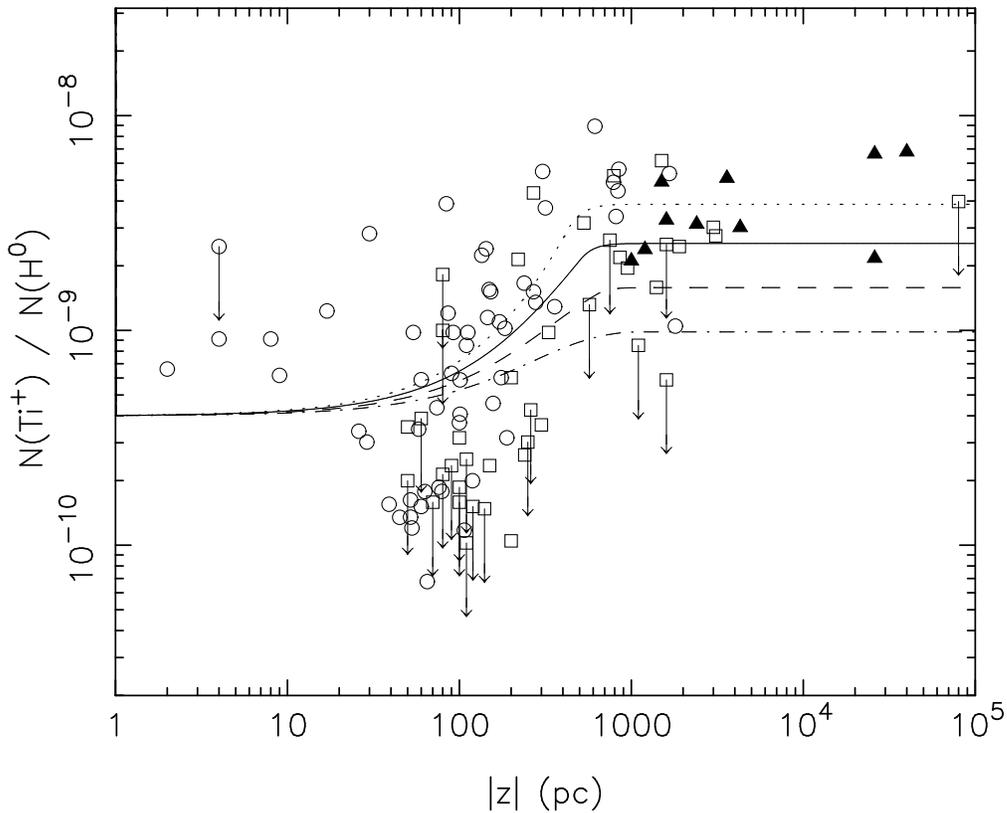

Fig. 3. — Values of Ti abundance measured towards 107 stars at distances $|z|$ from the Galactic plane. The symbols have the same meanings as in Figure 2. The four lines drawn give the predictions of a simple model for the density-dependence of the gas-phase abundance of Ti, discussed in the text, with power law index $k = -0.6$ (dot-dash line), $-0.8$ (long dash line), $-1.0$ (continuous line), and $-1.2$ (short dash line).

Typically, only 0.5% of the titanium is in gaseous form in the disk ISM. In contrast, both the average depletion and the scatter are reduced beyond $\mid z \mid \simeq 500$ pc; as these sight-lines also include material in the disk, the true values in the halo must be even lower. Although the number of measurements is not large, there seems to be little residual increase in the integrated $N(\mathrm{Ti}^+)/N(\mathrm{H}^0)$ over sight-lines which extend further than $\mid z \mid \simeq 500$ pc; the median value of the points at $\mid z \mid \geq 500$ pc corresponds to a titanium depletion $D(\mathrm{Ti}) = -1.5$ .

In Figure 4 the titanium abundance is plotted as a function of the average volume density of neutral gas along the line of sight, $N(\mathrm{H}^0)/d$, where $d$ is the distance to the star. Similar relationships for several other elements have been presented by Jenkins (1987). Again, the lines in Figure 4 refer to the model discussed below. The strong trend of increasing $A(\mathrm{Ti})_{ISM}$ with decreasing $< n(\mathrm{H}^0) >$ noted by earlier workers can be seen clearly in Figure 4. We deduce a gradient of $m = -0.70$, compared with Jenkins' value of $-0.84$ . [1]

### 4.3 Interpretation

We now attempt to bring together into one coherent picture the information conveyed by the data in Figures 2, 3, and 4. We adopt a simple model for the distribution of interstellar titanium with distance from the plane of the Galaxy, based on two assumptions:

(1) The $z-$profile of $\mathrm{H}^0$ is represented by the exponential fit

$$n_z(\mathrm{H}^0) = n_0(\mathrm{H}^0) \, \exp\left[-\mid z \mid / h_{H^0}\right] \tag{9}$$

with parameters

$$n_0(\mathrm{H}^0) = 0.78 \ \mathrm{cm}^{-3}; \ h_{H^0} = 95 \ \mathrm{pc} \tag{10}$$

appropriate to our sample (section 4.1 above); and

(2) the density dependence of the Ti abundance is a power law of the form

$$n(\mathrm{Ti}^+)/n(\mathrm{H}^0) = A_0 \times [n(\mathrm{H}^0)]^k \tag{11}$$

where $k$ is negative. Combining eqs. (9) and (11), we obtain:

$$n_z(\mathrm{Ti}^+) = A_0 \times \left[n_0(\mathrm{H}^0)\right]^{k+1} \, \exp\left[-\mid z \mid \times (k+1)/h_{H^0}\right] \tag{12}$$

for the gas-phase density of Ti as a function of distance $\mid z \mid$ from the plane.

---

[1] Earlier versions of this plot (e.g. Jenkins 1987; Albert et al. 1993) show an additional point near $< n(\mathrm{H}^0) > = 0.004$ cm$^{-3}$ and $N(\mathrm{Ti}^+)/N(\mathrm{H}^0) = 4 \times 10^{-8}$, which is clearly separated from the rest of the measurements. We are of the opinion that this point, which refers to observations by Stokes (1978) of the sight-line to $\eta$ U Ma, has been incorrectly represented. No absorption can be discerned in the original spectrum by Stokes, and the point should therefore be considered an upper limit rather than a detection.

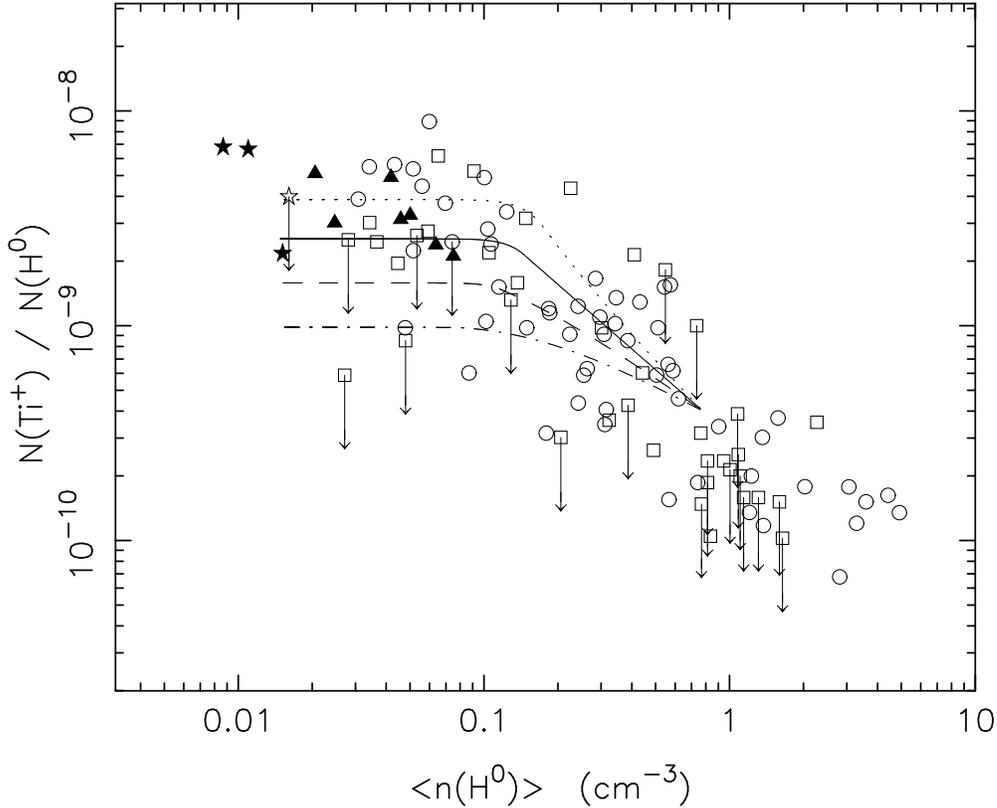

Fig. 4. — Ti abundance plotted as a function of the average density of neutral gas along the 107 sightlines surveyed. The symbols are as in Figure 2; the values of $< n(H^0) >$ for the four extragalactic sightlines (plotted with star symbols) assume that the Galactic halo extends to $| z | = 5$ kpc. The four lines drawn have the same meaning as in Figure 3; they originate at the median value of the Ti abundance $(A(Ti)_{ISM} = 4 \times 10^{-10})$ at the base of the halo, where the median gas density is $n_0(H^0) = 0.78$ cm$^{-3}$.

This model has two free parameters, $A_0$ and the power-law exponent $k$. By fixing $A_0 \times [n_0(H^0)]^k = 4 \times 10^{-10}$, the mid-plane value of the Ti abundance deduced from the median $N(Ti^+)/N(H^0)$ for stars within 100 pc of the plane, and by further imposing the boundary condition that $A(Ti)_{ISM} \leq A(Ti)_\odot$, we can check how different values of $k$ compare with the data in Figures 3 and 4 by integrating eq. (12). The lines drawn in these Figures correspond to $k = -0.6, -0.8, -1.0$, and $-1.2$ respectively; a reasonable fit is apparently obtained with $k \simeq -1$. In Figure 5 we show the predictions of the model superposed on the same data as in Figure 2; again a value of $k \simeq -1$ is suggested.

It is important to stress that the comparisons between observations and model predictions shown in Figures 3, 4, and 5 are only of a qualitative nature, because of the intrinsic scatter of the data and the uncertainties in the choice of model parameters. Nevertheless, the model does appear to reproduce broadly the overall trends in Figures 3, 4, and 5, and a rapid increase in Ti abundance with decreasing gas density above the plane seems indicated.

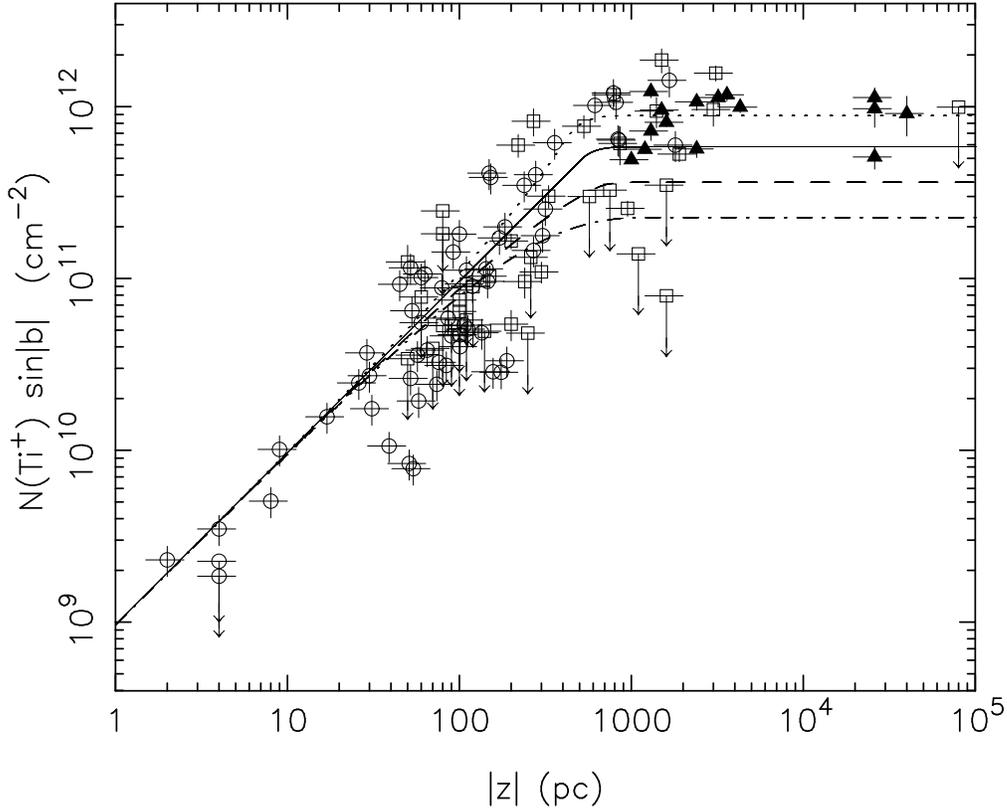

Fig. 5. — The observed distribution of Ti$^+$ column densities with distance from the Galactic plane is compared with those predicted by our simple model of the density-dependence of the Ti abundance. The data points are the same as in Figure 2; the four lines drawn are for power-law index $k = -0.6$ (dot-dash line), $-0.8$ (long dash line), $-1.0$ (continuous line), and $-1.2$ (short dash line).

Before discussing the implications of this result, we note that the approach described above differs from earlier analyses in one important aspect. Previous workers (e.g. Edgar & Savage 1989) have taken the value of $k$ to be the slope in Figure 4. However, that slope measures the ratio of *column densities* integrated along a given distance and is *not* the same as the index of the power law in eq. (11), which is a relationship between the *volume densities* of Ti$^+$ and H$^0$ .

## 5. CONCLUSIONS

We have presented observations of Ti II $\lambda3384$ absorption towards 15 distant stars in the Galactic halo and the Magellanic Clouds. These new data extend earlier surveys of interstellar titanium to larger distances from the disk than probed up to now. We report two principal findings:

(1) The scale height of $Ti^+$ in the Galaxy is $h_{Ti^+} = 1.5 \pm 0.2$ kpc, larger than those of other H I species but not extending to several kpc, as had been suspected. The large scale height probably reflects the reduced depletion of Ti in low density regions in the halo.

(2) A simple model in which the density dependence of the gas-phase abundance of Ti is a power law with exponent $k \simeq -1$ reproduces the observations adequately . If this interpretation is correct, the volume density of Ti is, *on average*, independent of the gas density (see eq. 11). Apparently, the release of Ti from dust balances overall the falling density of gas away from the Galactic plane, until all the titanium is in the gas phase at $|z| \sim 1$ kpc . This solution is not unique. In particular, a faster increase of $A(Ti)_{ISM}$ with decreasing $n(H^0)$ is admitted by the data if we restrict the maximum gas-phase abundance attainable to be *less* than the solar value, with some of the Ti remaining in solid form even at large distances from the disk.

It should be relatively straightforward in future to test the validity of our simple model by measuring the abundance of Ti directly in individual absorbing clouds which are further than $|z| \simeq 1$ kpc. Relevant data are at present available for only a few cases. The clouds at $v_{LSR} = +32$ and $+61$ km s$^{-1}$ in R 136 (see Figure 1) are located more than $\sim 5$ kpc below the plane, if their velocities reflect the overall rotation of the Galaxy rather than peculiar motions (Savage & de Boer 1981). By comparing our measured $N(Ti^+) = 6.5 \times 10^{11}$ and $3.5 \times 10^{11}$ cm$^{-2}$ (Table 3) with $N(H^0)_{21 cm} = 2.6 \times 10^{18}$ and $7.1 \times 10^{17}$ cm$^{-2}$ (McGee, Newton & Morton 1983), we deduce $D(Ti) = +0.5$ and $+0.8$ respectively (eq. 8). Within the uncertainties introduced by the 15 arcmin beamwidth of the 21 cm observations by McGee et al., these results are consistent with an approximately solar abundance of Ti in these two halo clouds. McGee et al. reached a similar conclusion for other refractory elements, such as Al, Si, and Fe, in the halo clouds at $v_{LSR} = +70$ and $+120$ km s$^{-1}$ towards HD 36402, also in the Large Magellanic Cloud. As a counter-example, in their comprehensive analysis of *HST* observations of the halo star HD 93521 Spitzer & Fitzpatrick (1993) found the release of Ti from the grains to be less pronounced than that of Fe, with residual depletions $D(Ti) = -0.9$ in clouds between $v_{LSR} = -39$ and $-66$ km s$^{-1}$.

Establishing if the solar abundances of refractory elements such as Ti are indeed recovered and little dust remains beyond $\sim 1$ kpc from the plane would provide valuable clues to the processes which maintain the interstellar medium in the halo. In particular, the relationship of this extended, neutral component to the hot gas producing C IV absorption and emission (see for example the review by Savage 1991) is still unclear. While the coronal gas presumably arises in a Galactic fountain fed by supernovae (Spitzer 1990; Shapiro & Benjamin 1992; Shull & Slavin 1994), there may be difficulties in applying the same ideas to explain the distribution of neutral gas within a few kpc of the disk (Houck & Bregman 1990).

It is a pleasure to acknowledge the help by Richard Hunstead and David Mar in obtaining some of the data on our behalf, particularly during a brief period in 1990 when one of us (M.P.) had to abstain from the joys of observing. The La Palma and AAT service programmes were very useful in allowing us to complete this project. We thank Geraint Lewis for his assistance with the computational aspects of this work, and Barry Welsh for kindly communicating recent results in advance of publication. We are very grateful to Blair Savage and Ken Sembach for their constructive comments on an earlier version of this paper.